\documentclass[epjCONF]{svjour}
\usepackage{graphics}
\usepackage[varg]{txfonts} 
\usepackage[latin1]{inputenc}
 \newcommand\la{\langle}
 \newcommand\ra{\rangle}
 \newcommand\beq{\begin{equation}}
 
 \newcommand\eeq{\end{equation}}
 \newcommand\beqn{\begin{eqnarray}}
 \newcommand\eeqn{\end{eqnarray}}
 \newcommand\nn{\nonumber}
\def\J{J/\Psi}

\def\fm{\,\mbox{fm}}

\def\GeV{\,\mbox{GeV}}
\def\TeV{\,\mbox{TeV}}

\session-title{International Conference on New Frontiers in Physics}
\begin{document}
\title{\boldmath$J/\Psi$ production off nuclei: a detour from SPS to LHC}
\author{B.Z. Kopeliovich\thanks{\email{boris.kopeliovich@usm.cl}} \and I.K. Potashnikova\thanks{\email{irina.potashnikova@usm.cl}} \and Iv\'an Schmidt\thanks{\email{ivan.schmidt@usm.cl}}}
\institute{Departamento de F\'{\i}sica, Universidad T\'ecnica Federico Santa Mar\'{\i}a,\\
Centro de Estudios
Subat\'omicos, and Centro Cient\'ifico-Tecnol\'ogico de Valpara\'iso,\\
Casilla 110-V, Valpara\'iso, Chile}
\abstract{We present a collection of selected phenomena observed  in $\J$ production from proton-nucleus and heavy ion collisions at energies, ranging between the SPS and LHC. The emphasis is placed on the related theoretical  
ideas or techniques, which are either not widely known, or offer an alternative explanation to the observed
nuclear effects.}

\maketitle

\section{Preface}
\label{intro}

Nuclear suppression of heavy quarkonia is usually considered
as a hard probe sensitive to the properties of the
short-living medium produced in heavy ion collisions. 
The main challenge is to discriminate between initial
state interactions (ISI), usually identified as cold nuclear
matter effects, and final state interaction (FSI) and attenuation of the produced quarkonium in the
dense matter created in the nuclear collision. While the
latter is the main goal of the study, the result  depends on how well can one 
single out FSI from the admixture of ISI, which cannot be measured, but only theoretically modeled.
Important information on the cold nuclear matter effects can be learned from data on proton-nucleus collisions.
However, there is no simple recipe for extrapolation of such information from $pA$ to ISI in $AA$ collisions. Here we identify several obstacles preventing one from doing that easily. The mechanisms of ISI considerably vary between the energies of SPS and LHC. 

\section{Cold nuclear matter: pA collisions}

\subsection{Evolution and absorption of a charm dipole in a medium}
\label{abs}

\begin{figure}[h!]
\centerline{
\resizebox{0.25\columnwidth}{!}{
\includegraphics{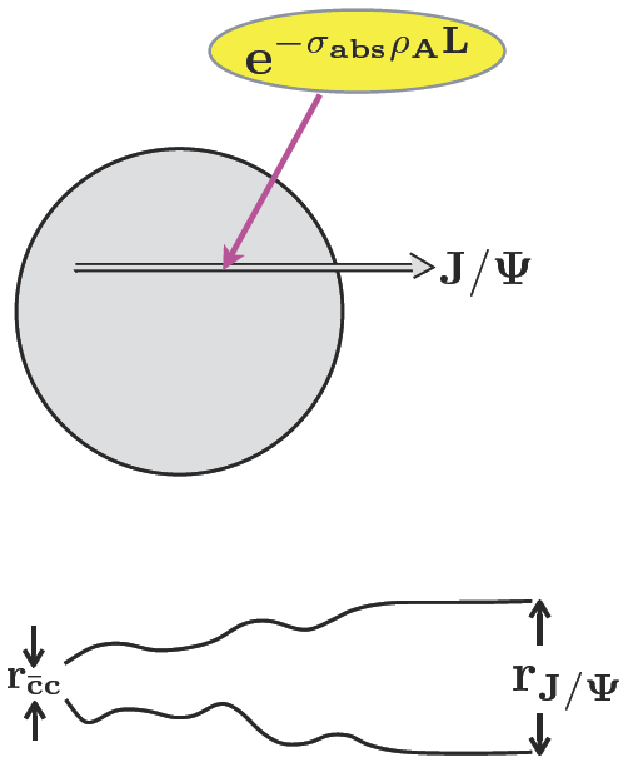} }\hspace{2cm}
\resizebox{0.35\columnwidth}{!}{
\includegraphics{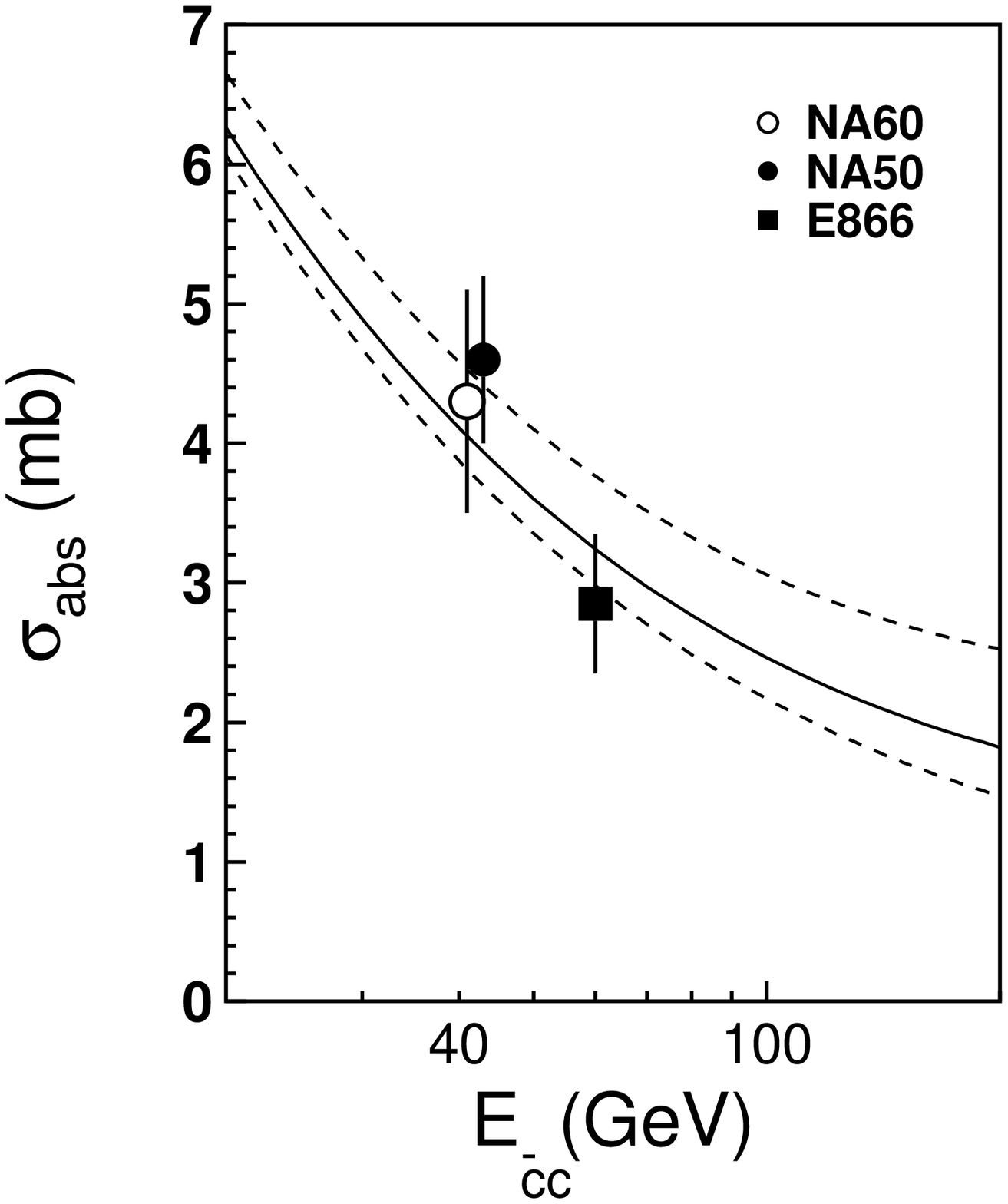}}}
\caption{{\it Left upper:} illustration for exponential attenuation of the produced charmonium.
{\it Left bottom:} illustration for evolution of a $\bar cc$ propagating though a medium.
{\it Right:} the break-up cross section fitted with expression (\ref{10}) to data from fixed target experiments \cite{na60}. The curves show the results of a similar fit to the theoretical calculations of the $\J$ production rate with 
Eq.~(\ref{80}) and $\delta=1/2,\ 1,\ 2$, from bottom to top respectively.
} 
\label{fig:sig-eff}      
\end{figure}

Usually the nuclear ratio is evaluated with an oversimplified model \cite{na60} assuming that $\J$ attenuates with a constant cross section $\sigma_{abs}$ on the way out of the nucleus, as is illustrated in Fig.~\ref{fig:sig-eff}
(left). Correspondingly, the nuclear modification factor has the form,
\beq
R_{pA}=\frac{1}{A\sigma_{abs}}\int d^2b\,\left[1-e^{-\sigma_{abs}T_A(b)}\right],
\label{10}
\eeq
where $\sigma_{abs}$ is treated as an unknown  parameter fitted to data; $T_A(b)=\int_{-\infty}^\infty
dz\,\rho_A(b,z)$ is the impact parameter dependent nuclear thickness function. The results of such an analysis at different energies of $\J$ plotted in Fig.~\ref{fig:sig-eff} (right), demonstrate a steep decrease of the effective absorption cross section with energy.

However, as is illustrated in the left bottom part of Fig.~\ref{fig:sig-eff}, a $\bar cc$  dipole is produced with a starting small separation $r_{\bar cc}\sim 1/m_c\sim 0.1\fm$, then is
evolving up to the $J/\Psi$ mean size $r_{J/\Psi}\sim 0.5\fm$ and eventually forms the wave function during the formation time 
\beq
t_f=\frac{2E_{\psi}}{m^2_{\Psi'}-m^2_{\psi}} = 0.1\fm\,\left(\frac{E_{\psi}}{1\GeV}\right)
\label{20}
\eeq

The expansion rate can be estimated perturbatively as,
\beq
\frac{dr_T}{dt}=\frac{4p_T}{E_{\bar cc}}.
\label{30}
\eeq
Employing the uncertainty relation one gets,
\beq
r_T^2(t)=\frac{8t}{E_{\bar cc}}+\frac{\delta}{m_c^2}.
\label{40}
\eeq
The dipole cross section of a $\bar cc$ dipole with small transverse separation $r_T$ and energy $E_{\bar cc}$
can be approximated as  \cite{zkl},
\beq
\sigma_{\bar cc}(E_{\bar cc})=C(E_{\bar cc})\,r_T^2.
\label{45}
\eeq
Then the absorption cross section as function of $\J$ energy and path-length takes the form,
\beq
\bar\sigma_{abs}(L,E_{\bar cc})={1\over L}\int\limits_0^L dl\,\sigma_{abs}(l) = C(E_{\bar cc})\,\left(\frac{4L}{E_{\bar cc}}+
\frac{\delta}{m_c^2}\right).
\label{50}
\eeq
With such a varying cross section one can easily calculate the nuclear modification ratio $R_{pA}$, compare with Eq.~(\ref{10}), and adjust the effective cross section. The results are shown in Fig.~\ref{fig:sig-eff} (right) by three curves 
corresponding to $\delta=2,\ 1,\ 0.5$ from top to bottom respectively. Similar analysis of data \cite{na60}
presented in Fig.~\ref{fig:sig-eff}
agree with these calculations.

Qualitatively, the reason for the observed falling with energy $\sigma_{abs}$ is clear: this is a manifestation of the color transparency effect \cite{zkl}. Indeed, the higher is the $\J$ energy, the more the initial small size of the $\bar cc$ dipole is ``frozen" by Lorentz time dilation, the more transparent the nuclear medium is. Notice that this can also
interpreted within hadronic representation as a multi-channel problem \cite{hk-prl}, which is equivalent, but technically more difficult description.

\subsubsection{pA: Higher twist c-quark shadowing}

At higher energies $\bar\sigma_{abs}$       is affected by another 
time scale, the lifetime of a   $g\to\bar cc$   fluctuation  inside the incoming proton, which can also be interpreted as a time scale for $\bar cc$ pair production,
\beq
t_p=\frac{2E_{\psi}}{m^2_{\psi}}=\frac{1}{x_2m_N},
\eeq
where $x_{1,2}$ are the usual Drell-Yan variables.
This time scale is about 5 times shorter than $t_f$.

\begin{figure}[h!]
\centerline{
\resizebox{0.65\columnwidth}{!}{
\includegraphics{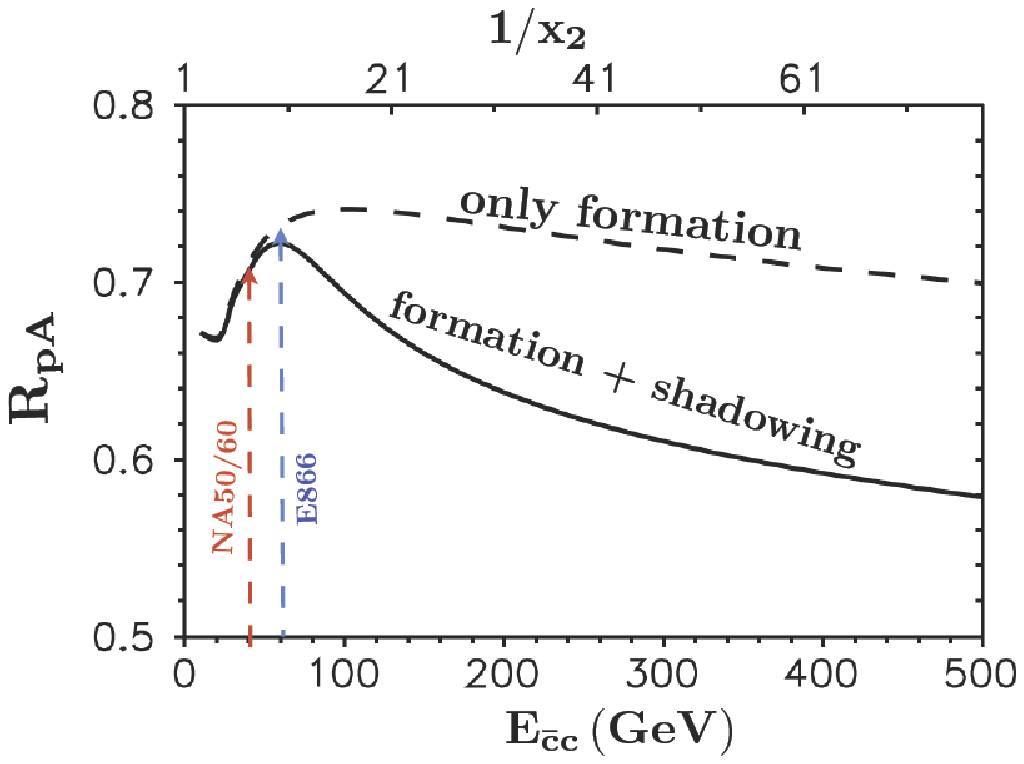}}\hspace{-0.5cm}
\resizebox{0.35\columnwidth}{!}{
\includegraphics{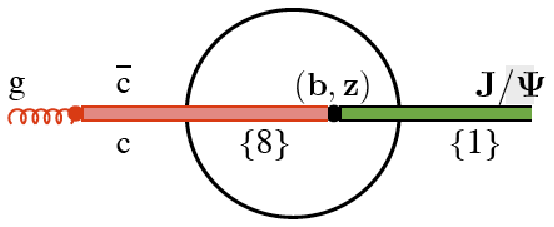} }}
\caption{{\it Left:} Nuclear ratio $R_{pA}$ for lead as function of $\bar cc$ energy, calculated with the path integral technique \cite{kth}. {\it Right:} Transition of the projectile gluon into a color octet dipole,  $g\to\{\bar cc\}_8$,
which propagates and attenuate in the nucleus, and interacts producing a color singlet dipole on a nucleon with coordinates $z,\vec b$. }
\label{fig:e-dep}      
\end{figure}

If $t_p \gtrsim R_A$       the initial state fluctuation $g\to\bar cc$ leads to shadowing corrections related to a non-zero    $\bar cc$ separation. This is a higher twist effect, which can be calculated within the dipole approach. 
In this case the evolution of a dipole should be treated with a more advanced theoretical tool, compared with the simple model used in the previous section.
Here we rely on the strict quantum-mechanical description the dipole evolution in a medium, the path-integral technique \cite{kz91}, which sums up all possible paths of the quarks. The evolution equation in the light-cone variables has a form of 
the Schr\"odinger equation for the Green function, describing propagation of a dipole from the initial longitudinal position $z_1$ and dipole size $\vec r_1$ up to $z_2$ and $\vec r_2$ \cite{kz91,kst2},
\beq
i\frac{d}{dz_2}\,G(\vec r_2,z_2;\vec r_1,z_1) =
\left[\frac{2}{E_\psi}\,(m_c^{2} - \Delta_{r_{2}})
+ V(\vec r_2,z_2)\right]
G (\vec r_2,z_2;\vec r_1,z_1).
\label{60}
\eeq
The real part of the light-cone potential is given by the binding $\bar cc$ potential, which is chosen in the oscillatory form, and the imaginary part is proportional to the dipole cross section, which is assumed to be proportional to $r_2^2$. 
With this conditions and realistic Woods-Saxon shape of the nuclear density, the calculated energy dependence of the nuclear modification factor \cite{kth} is plotted in Fig.~\ref{fig:e-dep} (left).
The primary rise of $R_{pA}$ occurs in the same energy range, which was presented in Fig.~\ref{fig:sig-eff}.
This rise is due to the color transparency effect, and it would continue as is depicted by the dashed curve, if no other effects were contributing. However, the charm shadowing effect, illustrated in Fig.~\ref{fig:e-dep} (right), which onsets at long production time Eq.~(\ref{50}), causes an additional sizable suppression, and the total result is plotted by solid curve.

\subsubsection {pA: Charmonium suppression at RHIC/LHC}

At the energies  of RHIC and LHC all coherence time scales become long, $t_f>t_p\gg R_A$,
so the Green function Eq.~(\ref{60}) approaches the asymptotic limit of a "frozen" dipole size, $G (\vec r_2,z_2;\vec r_1,z_1)=\delta(\vec r_1-\vec r_2)$ for $z_2-z_1\sim R_A$. In this case the path integral formalism is essentially simplified, and the nuclear modification factor at a given impact parameter and longitudinal coordinate $z$ of the color exchange interaction, takes the form \cite{kth},
\beq
R_{pA}(b,z)=\int d^2r_T\,
K_0(m_c r_T)\,r_T^2\,\Psi_{\psi}(r_T)
e^{-{1\over2}\sigma_{\bar ccg}(r_T)T_-(b,z)
-{1\over2}\sigma_{\bar cc}(r_T)T_+(b,z)},
\label{70}
\eeq
where $T_-(b,z)=\int_{-\infty}^z dz'\rho_A(b,z')$; 
$T_+(b,z)=T_A(b)-T_-(b,z)$;
$T_A(b)=T_-(b,\infty)$; and
\beq
\sigma_{\bar ccg}(r_T)={9\over4}\sigma_{\bar cc}(r_T/2)-{1\over8}\sigma_{\bar cc}(r_T)
\label{80}
\eeq
Naively, one could think that a color octet $\bar cc$ pair propagating through the nucleus (see Fig.~\ref{fig:e-dep}, right) can experiences color exchanges remaining in the color octet states. So it cannot be absorbed, i.e. does not attenuate, and should not lead to initial-state shadowing. However, every process with a nonzero cross section
shadows itself with this cross section. The cross section of $\bar cc$ production is given by the three-body dipole cross section $\sigma_{\bar ccg}$ \cite{npz}, this is why it enters the exponent in Eq.~(\ref{70}) \cite{kth}.
Both cross sections $\sigma_{\bar ccg}$      and  $\sigma_{\bar cc}$    steeply
rise with rapidity $\sigma_{\bar cc} \propto Q_s^2(x_2)\propto e^{0.288\eta}$,
according to the parametrization \cite{gbw} fitted to DIS data from HERA.

\begin{figure}[h!]
\centerline{
\resizebox{0.48\columnwidth}{!}{
\includegraphics{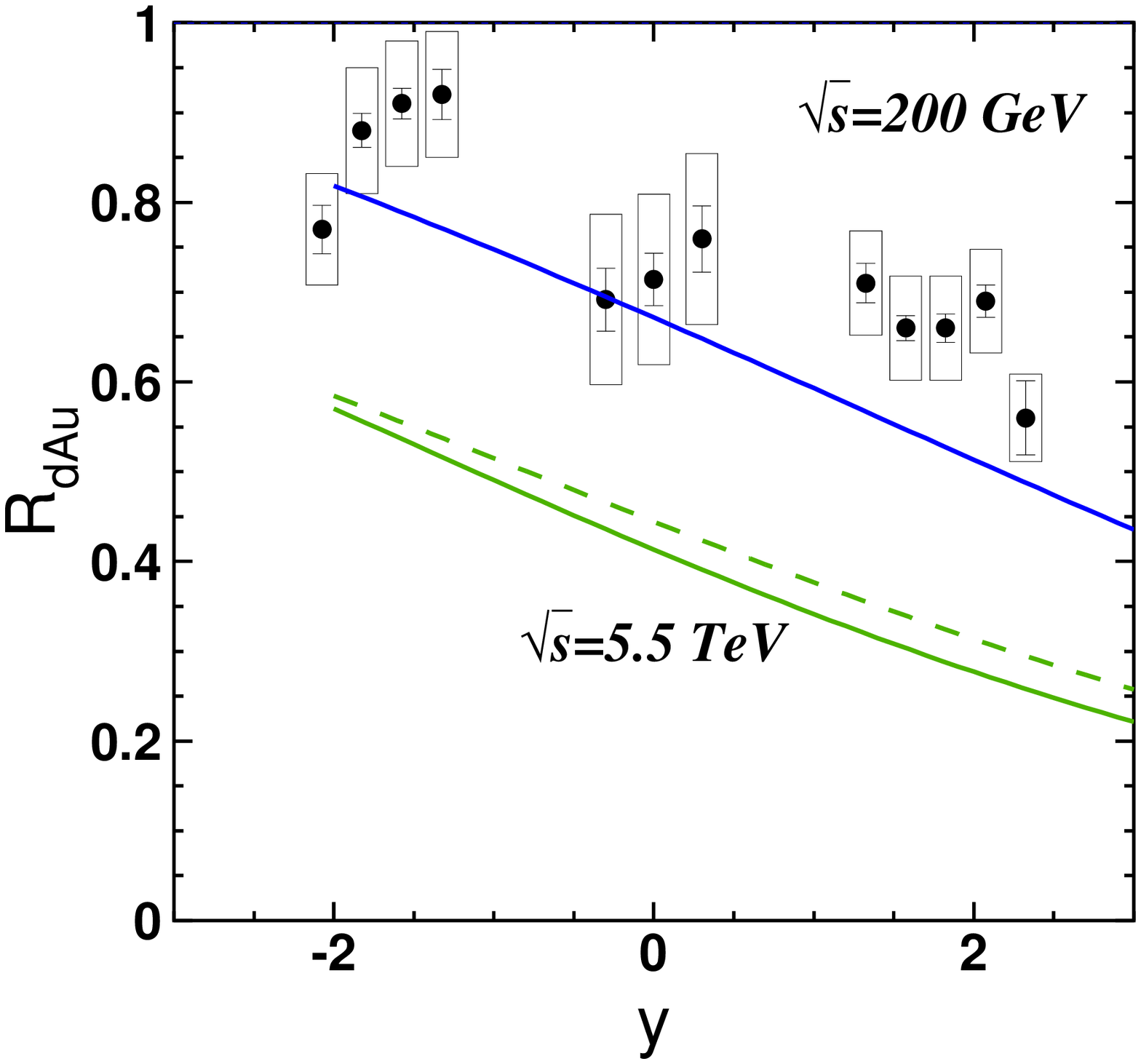} }\hspace{0cm}
\resizebox{0.48\columnwidth}{!}{
\includegraphics{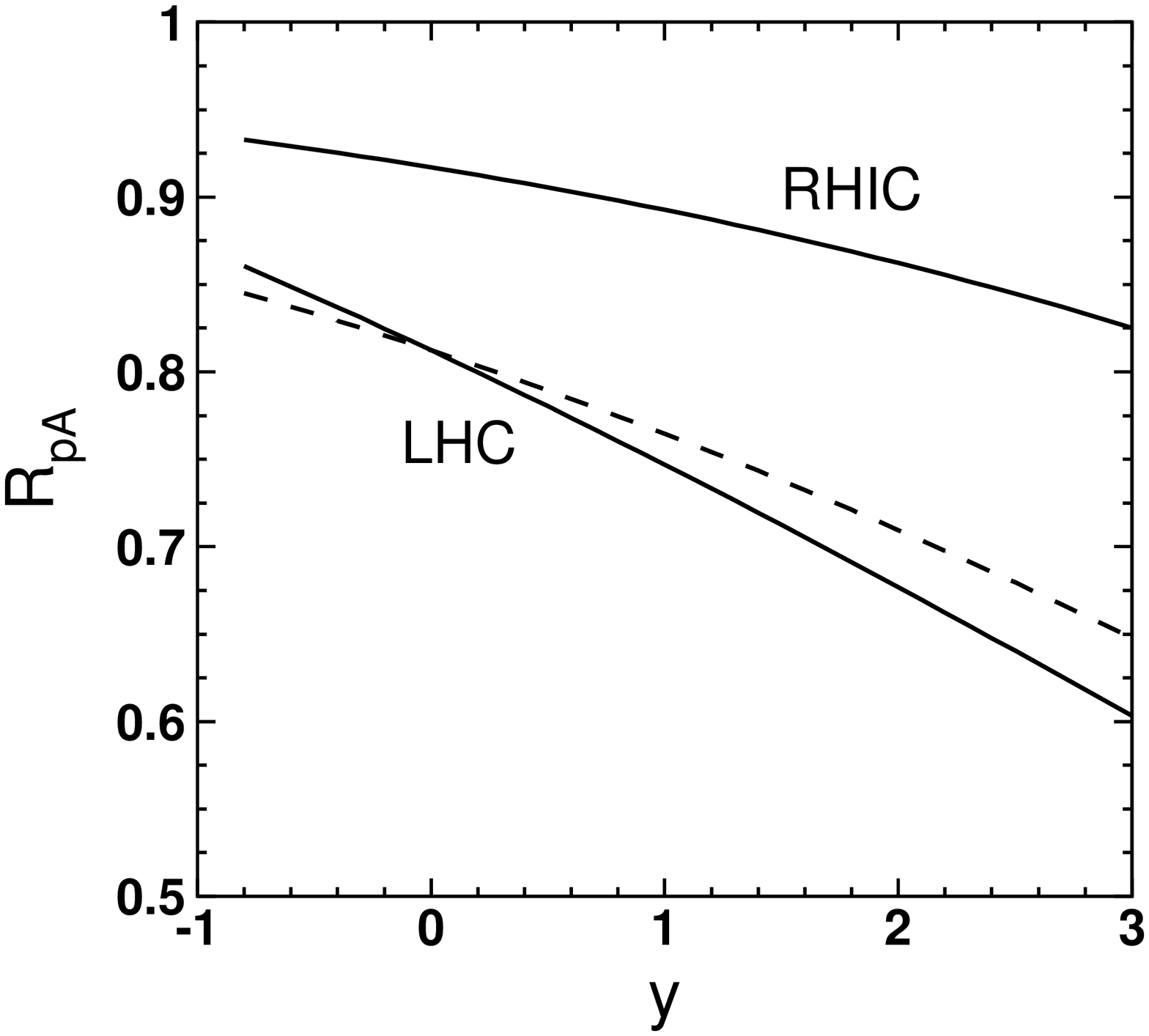}}}
\caption{{\it Left:} Data \cite{phenix-last} for the $p_T$-integrated nuclear suppression factor $R_{dAu}(y)$   for $\J$ produced in $d$-$Au$ collisions with rapidity $y$ at $\sqrt{s}=200\GeV$. The upper solid curve presents the results of calculations \cite{psi-lhc} at $\sqrt{s}=200\GeV$. Gluon shadowing is negligibly weak \cite{nontrivial,kst2,ds}. 
The lower solid (dashed) curve shows predictions for  proton-lead collisions at 
$\sqrt{s}=5.5\TeV$, including (excluding) gluon shadowing.
{\it Right:} All curves have the same meaning as in the left panel, but for $\Upsilon$ production.
} 
\label{y-dep}      
\end{figure}

The results of calculations are compared with RHIC data \cite{phenix-last} at $\sqrt{s}=200\GeV$ in the left panel of Fig.~\ref{y-dep}. 
The parameter free calculation \cite{psi-lhc} well agree with the data. The gluon shadowing correction was found to be negligibly small within the shown range of rapidity, because the gluon radiation coherence length is too short. This is confirmed by the global analysis \cite{ds} and dipole model calculations \cite{kst2}. 
Notice that these data are frequently described within a kind of 
an "upside-down" scenario. Namely, the break-up dipole cross section, which depends on energy and is well known from HERA data, is assumed to be an unknown constant. However, the magnitude of gluon shadowing, which is currently a controversial and model dependent issue \cite{kn-review}, is assumed to be known. 

Predictions for $\J$ suppression at the LHC energy $\sqrt{s}=5.5\TeV$ is also shown in the same plot. The dashed curve includes initial-state $c$-quark shadowing and final-state dipole break-up, given by Eq.~(\ref{70}). Solid curve differs by inclusion of gluon shadowing, which is visible, but still weak. 

Analogous mechanisms of nuclear suppression of heavy quarkonium production are applicable also to $\Upsilon$.
Suppression of the radial excitations       $\Upsilon(2S),\,\Upsilon(3S)$        
              is expected to be similar to $\Upsilon$ (compare  with $\Psi^\prime$ vs $J/\Psi$), since it is mainly
controlled by the size of the produced heavy dipole.
Our predictions for the energies of RIHC and LHC are plotted in the right panel of Fig.~\ref{y-dep}. 

\subsubsection{pA: why nuclear suppression scales in {\boldmath$x_F$}}

It has been observed in fixed target experiments \cite{na3,e866-psi} that nuclear suppression of $\J$ produced at forward rapidities in $pA$  collisions are strongly suppressed. Moreover, the $x_F$ dependence of the nuclear suppression factor is nearly independent of energy. Since large $x_F$ correspond to small $x_2$ of the target gluons, it is tempting to relate the increase of nuclear suppression with the coherence effects, like gluon shadowing \cite{kth}.
In this case, however, one would expect the suppression factor to scale in $x_2$, rather than $x_F$. Remarkably, the an enhanced suppression of particles produced at forward rapidities has been observed in any process, hard or soft, studied experimentally.
A natural explanation was proposed in \cite{brahms}.

Multiple interactions of the projectile hadron and its debris propagating through the nucleus should 
cause a dissipation of energy, what should result in a deficit of energy in a nuclear process at large $x_F$. This intuitive expectation is supported by consideration of the Fock 
state decomposition. The projectile hadron can be expanded over different states which are the 
fluctuations of this hadron.  In the limit of infinite momentum frame those fluctuations live 
forever. One can probe the Fock state expansion by interaction with a target. The interaction 
modifies the weights of the Fock states, a nuclear target enhances higher Fock components in the projectile hadron.
In each Fock component the hadron momentum is shared by the constituents, and the momentum 
distribution depends on their multiplicity: the more constituents are involved, the smaller is the 
mean energy per a constituent parton, i.e. the softer is the fractional energy distribution of a 
leading parton.
So on a nuclear target the projectile parton distribution  falls at {$ x_F\to1$} steeper than on a 
proton.  Apparently, this effect scales in $x_F$.
Further details of this mechanism of suppression and numerical results can be found in \cite{brahms,kn-review}.

\subsection{pA: Cronin effect}

Nuclear targets modify the transverse momentum distribution of produced particles, suppress it at small but enhance at medium large $p_T$. This effect named after Cronin, can be
calculated within the dipole model \cite{cronin,DY}. A simple description of the $p_T$-dependent nuclear modification factor for $\J$ production was proposed in \cite{psi-AA}. Available data on $\J$ produced in $pp$ collisions are well described by the following parametrization of the $p_T$-dependence ($p_T<5\GeV$),
\beq
\frac{d\sigma_{pp}(J/\Psi)}{dp_T^2}\propto
\left(1+\frac{p_T^2}{6\la p_T^2\ra}\right)^{-6}.
\label{90}
\eeq
 Following \cite{psi-AA,psi-bnl}, the simple way to calculate the nuclear modification factor is to make a shift in the mean square of $\J$ transverse momentum on the nuclear target $\la p_T^2\ra_{pA}=\la p_T^2\ra_{pp}+\Delta_{pA}$.
\beq
R_{pA}(p_T,b)=\frac{\la p_T^2\ra\,R_{pA}}{\la p_T^2\ra+\Delta_{pA}(b)}
\left(1+\frac{p_T^2}{6\la p_T^2\ra}\right)^6
\left(1+\frac{p_T^2}{6[\la p_T^2\ra+\Delta_{pA}(b)]}\right)^{-6}.
\label{100}
\eeq
Here $\Delta_{pA}(b)=\la p_T^2\ra_{pA}-\la p_T^2\ra_{pp}$ is the broadening for a $J/\Psi$ produced at impact parameter $b$. It was calculated in \cite{jkt} within the dipole approach as,
\beq
\Delta_{pA}(b)={9\over8}\ \bar\nabla_{\!\!r_T}^2\sigma_{dip}(r_T)\Bigr|_{r_T=0}\ T_A(b),
\label{110}
\eeq
in good agreement with available data for $J/\Psi$ broadening \cite{broadening,lanl}.

\begin{figure}[h!]
\centerline{
\resizebox{0.36\columnwidth}{!}{
\includegraphics{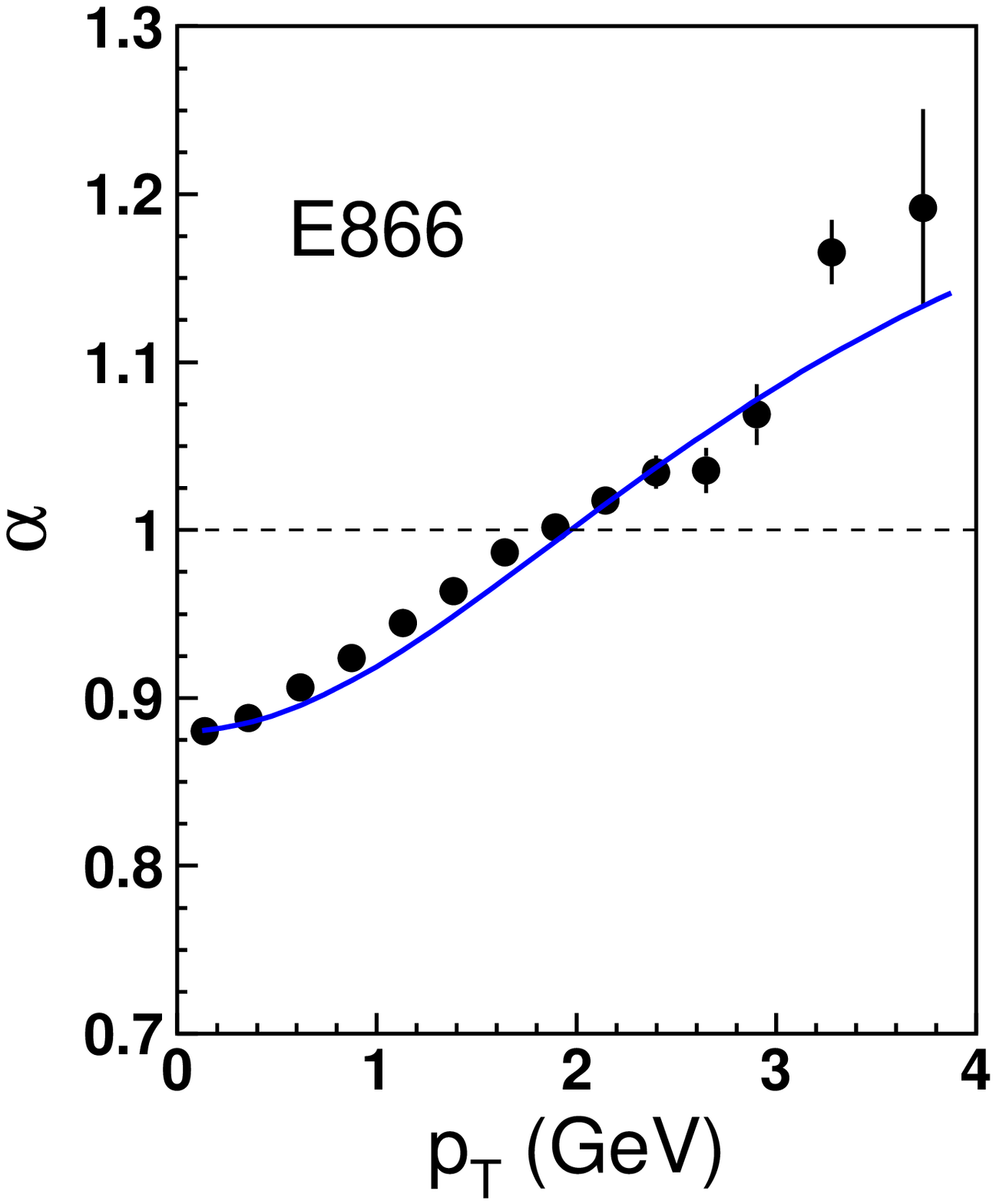} }\hspace{1cm}
\resizebox{0.49\columnwidth}{!}{
\includegraphics{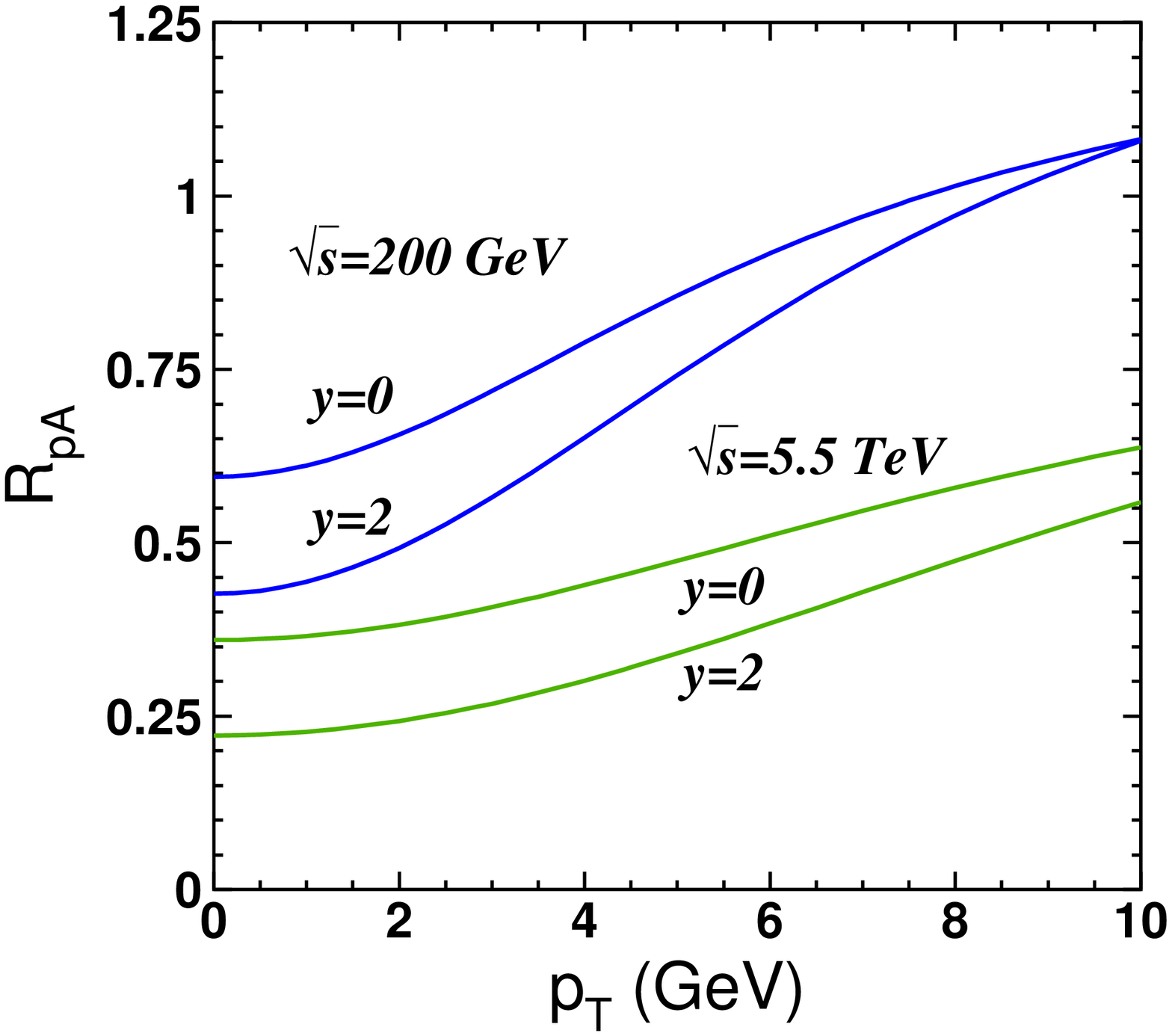}}}
\caption{{\it Left:} The exponent $\alpha=1+\ln(R_{pA})/\ln(A)$ as function of $p_T$ calculated with Eq.~(\ref{100}) in comparison with data from the E866 experiment \cite{e866}.
{\it Right:} Prediction for the $p_T$-distributions of $J/\Psi$ produced with rapidities $y=0,2$ in $p$-$Au$ collisions at $\sqrt{s}=200\GeV$ and in $p$-$Pb$ collisions at $\sqrt{s}=5.5\TeV$.
} 
\label{fig:pt-dep}      
\end{figure}

Formula (\ref{100}) is compared with data \cite{e866} in the left pane of Fig.~\ref{fig:pt-dep}, and predictions for RHIC and LHC are presented in the right pane. For the energy dependence of we use the parametrization from \cite{psi-pp}, $\la p_T^2\ra=[-2.4+0.6\,\ln{s}]\GeV^2$.

\section{ Initial state interactions (ISI): transition from pA to AA}

The nuclear modification of $\J$ production in $AA$ collisions originate from  the initial (ISI) and final state interaction (FSI) stages. The former includes interactions during the propagation of the nuclei through each other, while the latter 
corresponds to the interaction with the created matter, which occurs at a much longer time scale, when the high-energy parts of the nuclear debris are already far apart. 
Such a factorization into the two stages is possible, because no interference between them is possible.

We consider here several important aspects of ISI. Although it is tempting to extrapolate our experience with $pA$ interactions to ISI in $AA$ collisions, such a procedure is not as straightforward as it looks at the first glance \cite{nontrivial}. In particular the so called ``cold nuclear matter" in $AA$ collisions, turns out to be not cold at all.

\subsection{Broadening of \boldmath$J/\Psi$ in pA and AA collisions}\label{broad}

\begin{figure}[h!]
\centerline{
\resizebox{0.40\columnwidth}{!}{
\includegraphics{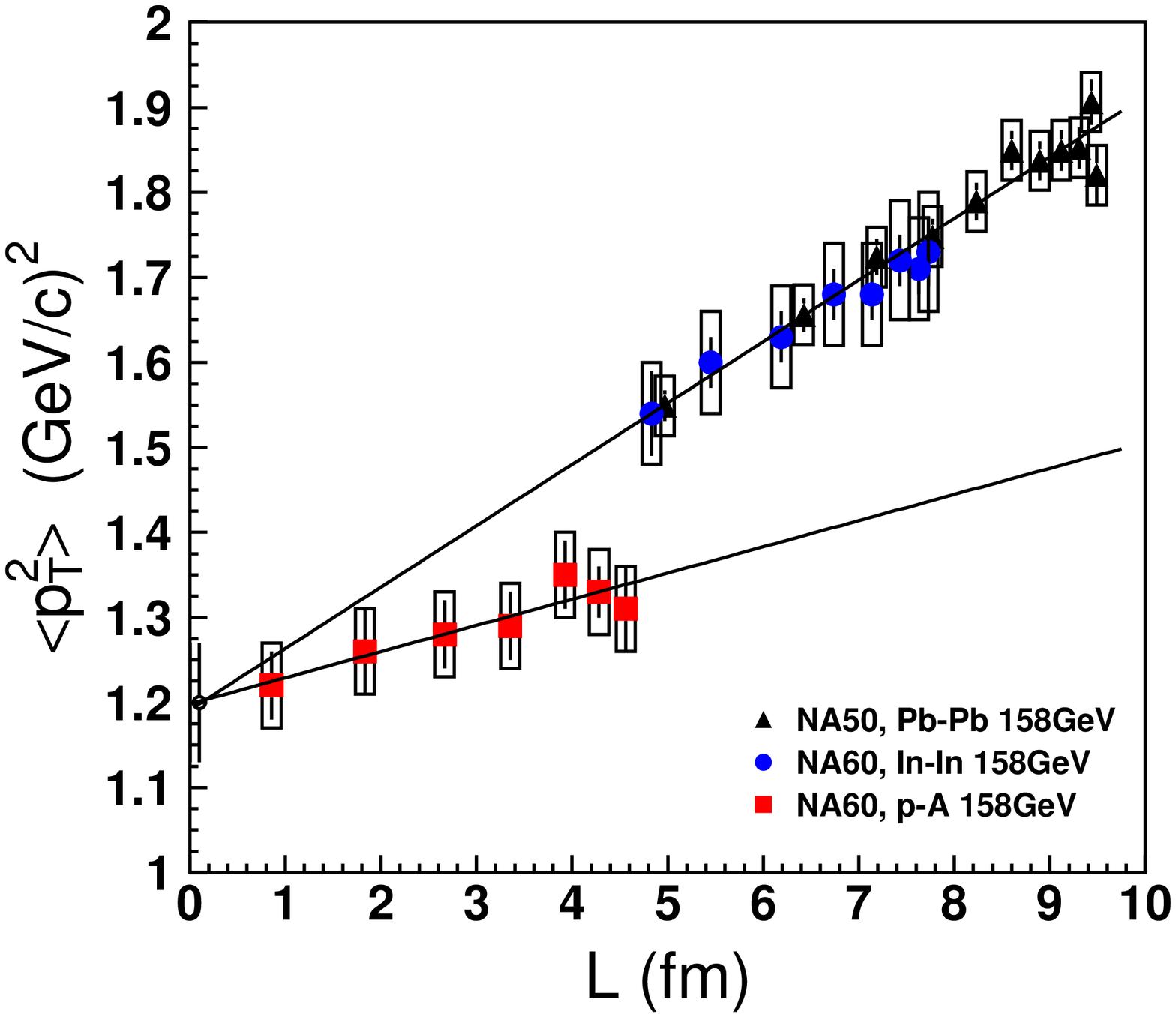} }\hspace{2cm}
\resizebox{0.30\columnwidth}{!}{
\includegraphics{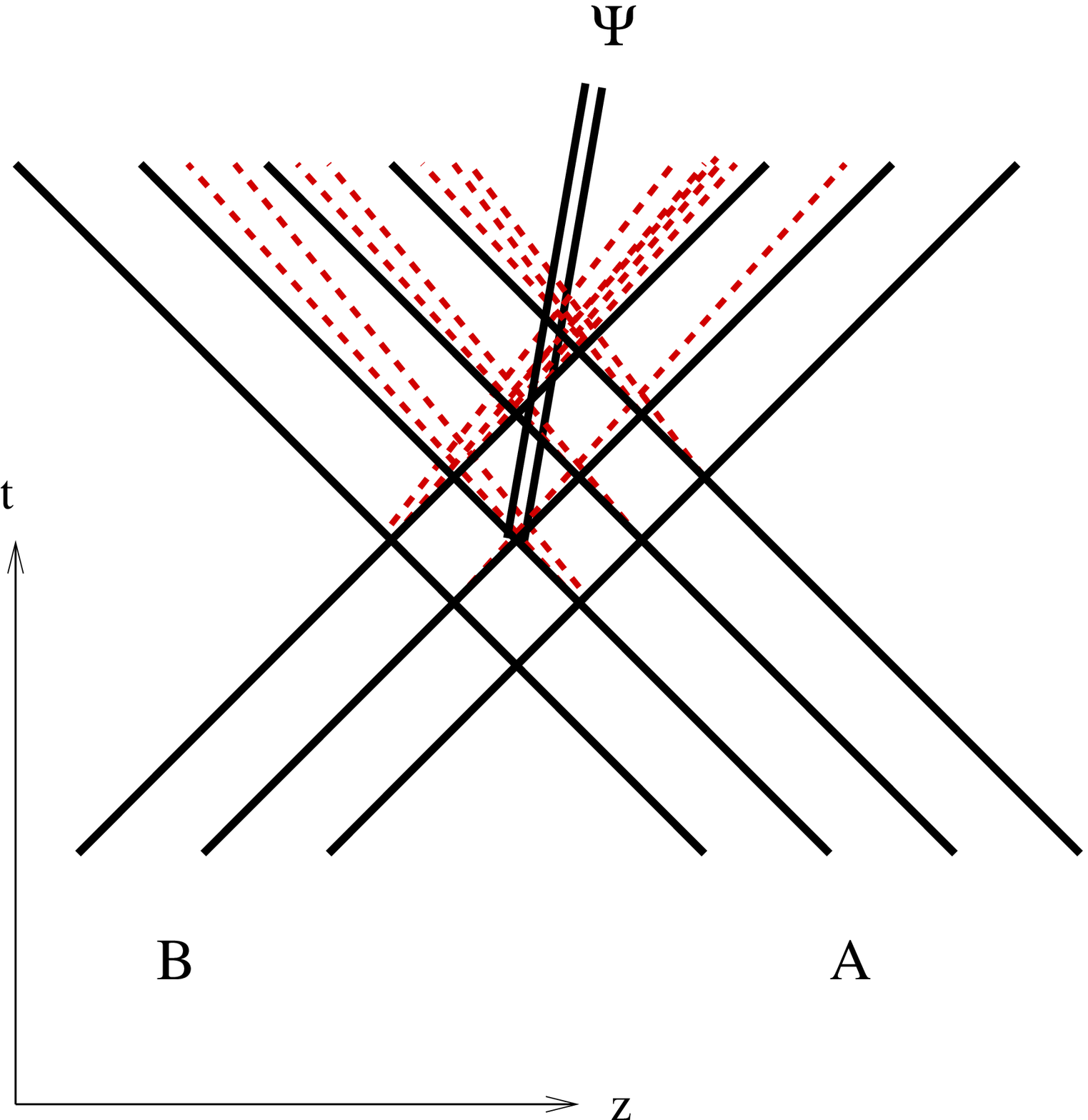}}}
\caption{{\it Left:} The mean transverse momentum squared of $J/\Psi$ produced in $pA$ and $AA$ collisions as function of the mean path length in the nuclear medium. Data points are from fixed target experiments \cite{na60-pt} at $158\GeV$.
{\it Right:} time - longitudinal coordinate plot for
charmonium production in a nuclear $AB$ collision in the c.m. The solid and dashed lines correspond to the nucleons and radiated gluons trajectories, respectively.
} 
\label{fig:broad}      
\end{figure}

Broadening is predominantly an ISI effect, since it is not affected by FSI\footnote{The cross section of $\J$ absorption in a medium is expected to rise with energy, i.e. with $p_T$, what should lead to shrinkage of  the $p_T$ distribution. We found this correction rather small.}.
This is a sensitive and unbiased probe for the properties on the ``cold nuclear matter". Indeed according to Eq.~(\ref{110}) broadening is proportional to the medium density integrated along the parton path length.
Thus, one should expect an universal broadening effect in $pA$ and $AA$ collisions, provided that the total path lengths
are equal. However the comparison done recently in \cite{na60-pt} and presented in Fig.~\ref{fig:broad} does not confirm such an universality.
Data show that broadening in nuclear collisions is about twice as large as in $pA$. This is a clear evidence of a significant difference between what is called ``cold nuclear matter" in $AA$ collisions and that in $pA$.

In fact, such a difference was predicted in \cite{hk,hhk}, where it was noticed that both the projectile gluon and the produced $\J$ (or the $\bar cc$ dipole) interact with bound "cold" nucleons in the case of a $pA$ collision, as is illustrated in the left pane of Fig.~\ref{fig:sig-eff}. However, in $AA$ collisions those target nucleons have already had a chance to interact with other nucleons in the beam nucleus, so they are the debris of the colliding nucleons and are not "cold" any more. Apparently $\J$ should interact with this nucleon debris with an increased cross section. In particular, such excited nucleons should be accompanied by radiated (on-mass-shell)
gluons, as is illustrated in the right pane of Fig.~\ref{fig:broad}.

The radiated gluons participate in the  $J/\Psi$  break-up,  as well as in broadening. Each on-mass-shell gluon
contributes as $\sigma _{\mathrm{abs}}^{\Psi g}\simeq \frac{9}{4}\sigma _{\mathrm{abs}}^{\Psi q}\simeq \frac{3}{4}\sigma _{\mathrm{abs}}^{\Psi N}$.
The mean number of radiated gluon per nucleon, $\la n_g\ra$, can be estimated relying on the time scale for 
gluon radiation, $l^g_f =2\,E_q\,x(1-x)/(x^2m_q^2 + k^2)$, where $x$ is the fractional momentum of the radiated gluon.
\beq
\la n_g\ra =\frac{3}{\sigma_{in}(NN)}\, \int\limits_{k^2_{min}}^{\infty}
dk^2 \int\limits_{x_{min}}^1 dx\, \frac{d\sigma(qN\to gX)}
{d\alpha\,dk^2}\, \Theta(\Delta z - l^g_f)
= \ \left\{\begin{array}{cc}6.9\times 10^{-1}\ \ &(\sqrt{s}=20\,GeV)\\ 
6.9\times 10^{-3}\ \ &(\sqrt{s}=200\,GeV)\\ 1.2\times
10^{-3}\ \ &(\sqrt{s}=1200\,GeV)\end{array}\right.
\label{130}
\eeq
We see that at the energy of SPS every participating nucleon has about one extra gluon, increasing its interaction cross section. However, the amount of such gluons steeply decreases with the collision energy and
vanishes at the energies of RHIC and LHC.

\begin{figure}[h!]
\centerline{
\resizebox{0.5\columnwidth}{!}{
\includegraphics{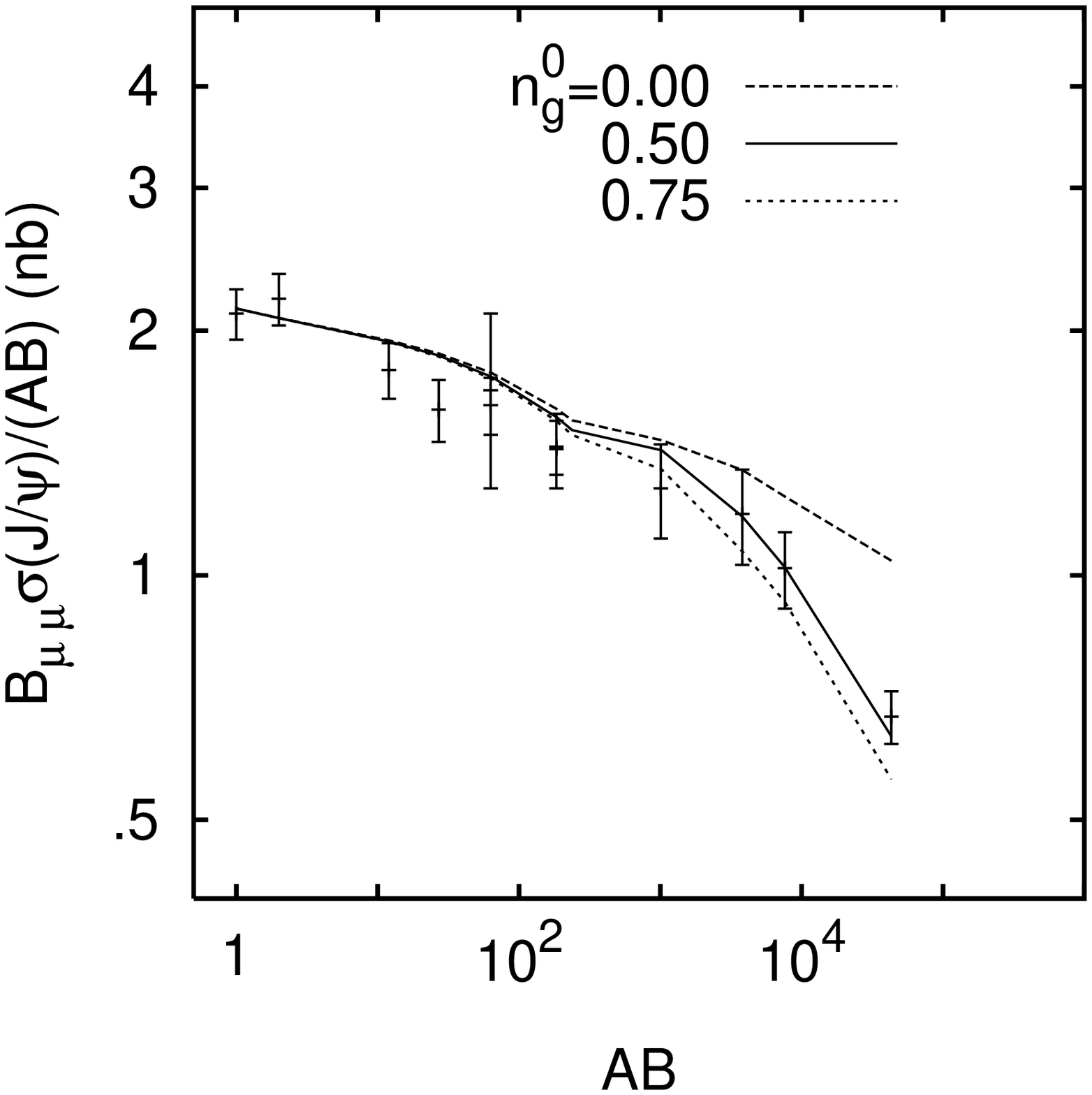} }\hspace{0cm}
\resizebox{0.5\columnwidth}{!}{
\includegraphics{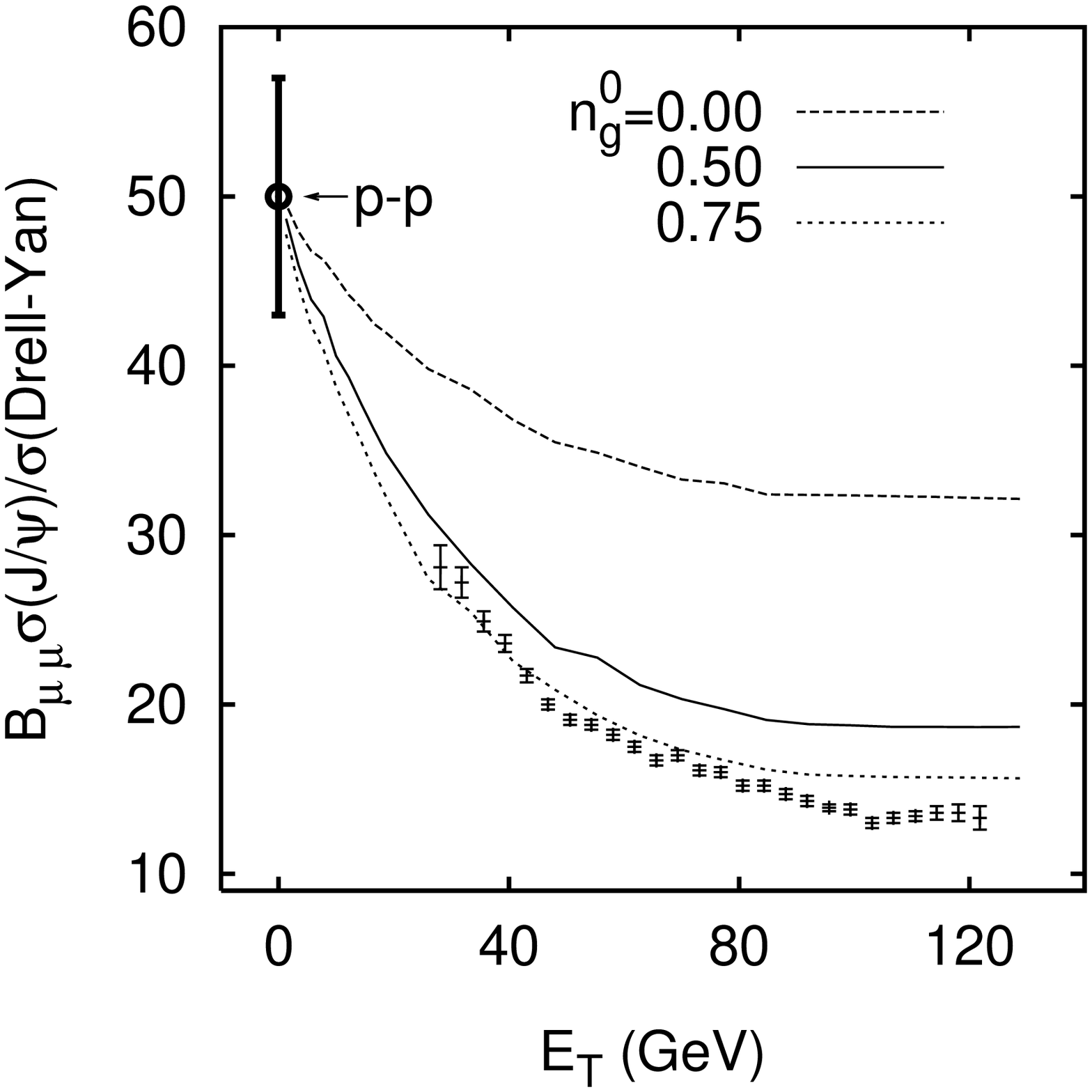}}}
\caption{{\it Left:} Ratio of $J/\psi$ to Drell-Yan cross section as a function
of centrality in S-U collisions at 200 GeV/c, with $\la n_{g}\ra=0,\, 0.5,\,
0.75$.
The data points are from \cite{NA3899}. The calculated curves
are normalized at $E_{T}=0$ to the ratio observed for p-p
collisions.
{\it Right:} The $ J/\psi$ total cross section as a function of the
product $AB$ of the projectile and target atomic mass numbers
at 200 GeV/c, for $\la n_{g}\ra=0,\, 0.5,\, 0.75$. The data
are from \cite{NA5097}} 
\label{fig:sps}      
\end{figure}

The results of calculations \cite{hk,hhk} are compared Fig.~\ref{fig:sps} with data on $\J$ suppression in minimum biased events (left) and as function of centrality (right). Apparently, the results with no radiated gluons 
grossly overestimate data for heavy nuclei and for central collisions, what lead to a conclusion about anomalous $\J$ suppression. At the same time, the data are well explained
with $\la n_g\ra\sim1$.
These result point at a possibility that the so called "anomalous" suppression of $\J$ observed in heavy ion collisions at the SPS, does not signal about a hot matter produced in final state, but is mainly a result of enhanced ISI.
The used as a baseline simple model assuming that ISI interactions of $\J$ are the same as in $pA$ collisions, is
just incorrect. This also explains why no jet quenching has been observed at SPS \cite{wang}.

Increase of the dipole-$N^*$ cross section by factor $1+0.75\la n_g\ra\approx1.75$ also well agrees with the observed larger broadening of $\J$ in $AA$ compared with $pA$ collisions, as is depicted in  Fig.~\ref{fig:broad} (left). Since FSI does not affect broadening, the observed strong increase confirms that
both "anomalous" effects of enhanced $\J$ suppression and broadening originate from ISI.

\subsection{Mutual boosting of the saturation scales in AA collisions}

The partonic structure of a hadron is known to depend on the hardness of the probe, the higher is the resolution, the more partons at small $x$ (and the less partons at $x\to1$) is resolved. This is controlled by the DGLAP evolution.
A nuclear target provides a harder probe for the partonic structure of the beam hadron compared with a proton target, because partons propagating through the nuclear target get an additional transverse kick, known as broadening.  Therefore, the projectile proton in $pA$ collisions acquires more partons at small $x$  than in $pp$ interactions.
In the case of $AA$ collisions all the participating nucleons in both nuclei change their partonic structure acquiring more low-$x$ gluons. This leads to an increase of the interaction cross section for such nucleons, resulting in enhanced broadening, which in turn excites the partonic structure of nucleons from another side even more. Such a mutual boosting
of parton density in the colliding nuclei leads to a rise of the saturation scales in the colliding nuclei, compared with $pA$ collisions. This process is described by the following bootstrap equations \cite{boosting}
\beqn
\tilde Q_{sB}^2(x_B)&=&\frac{3\pi^2}{2}\alpha_s(\tilde Q_{sA}^2+Q_0^2)
x_B g_N(x_B,\tilde Q_{sA}^2+Q_0^2)\,T_B\\ \nn
\tilde Q_{sA}^2(x_A)&=&\frac{3\pi^2}{2}\alpha_s(\tilde Q_{sB}^2+Q_0^2)
x_A g_N(x_A,\tilde Q_{sB}^2+Q_0^2)\,T_A.
\label{140}
\eeqn
Here $\tilde Q_{sA}^2(x_A)$ is the boosted saturation scale in the nucleus $A$; $Q_0^2$ is an infrared cutoff providing
the correct behavior in the soft limit (see in \cite{boosting}); $T_{A,B}$ are the nuclear thickness functions of the colliding nuclei. 

\begin{figure}[h!]
\centerline{
\resizebox{0.4\columnwidth}{!}{
\includegraphics{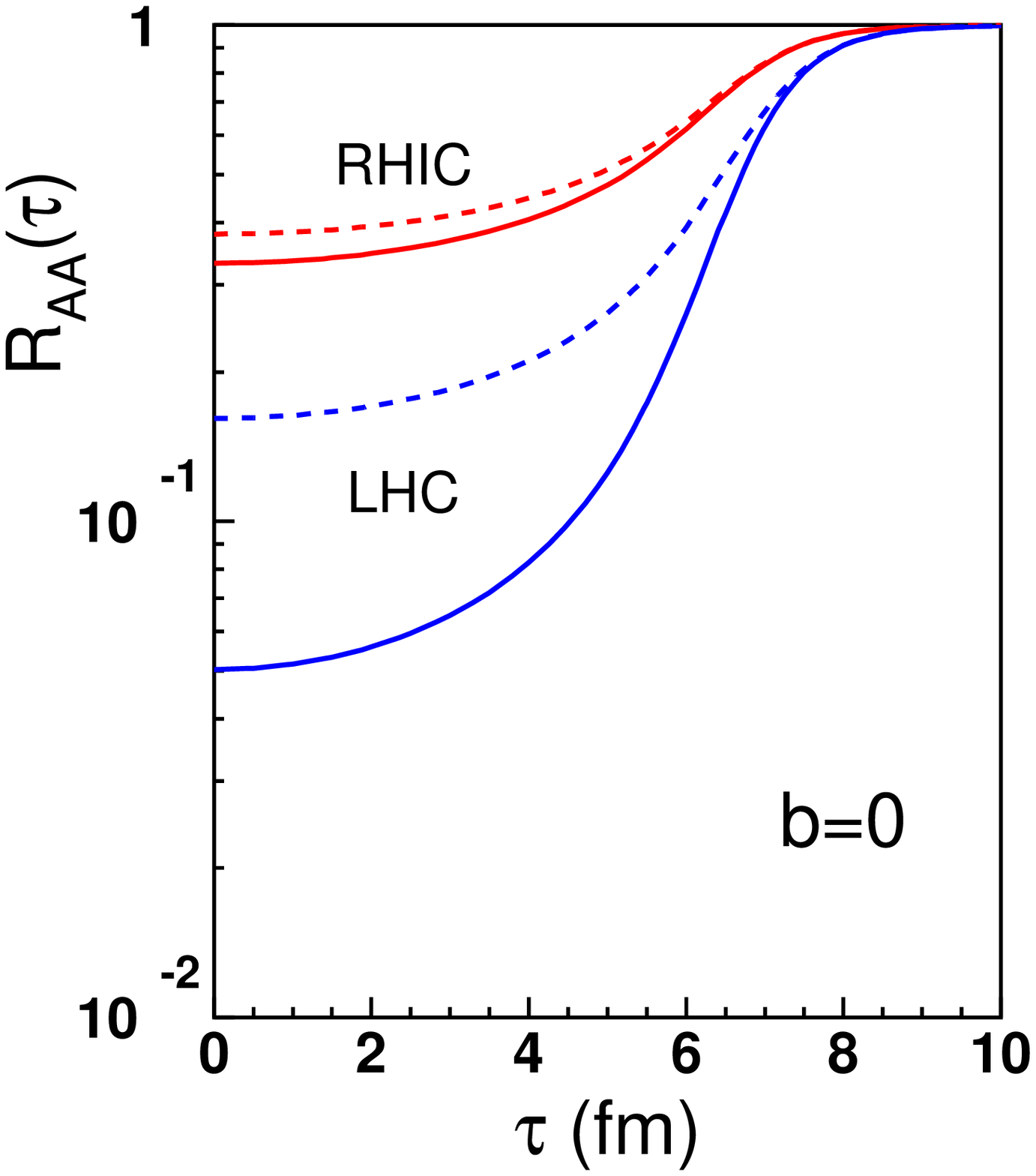} }\hspace{1cm}
\resizebox{0.4\columnwidth}{!}{
\includegraphics{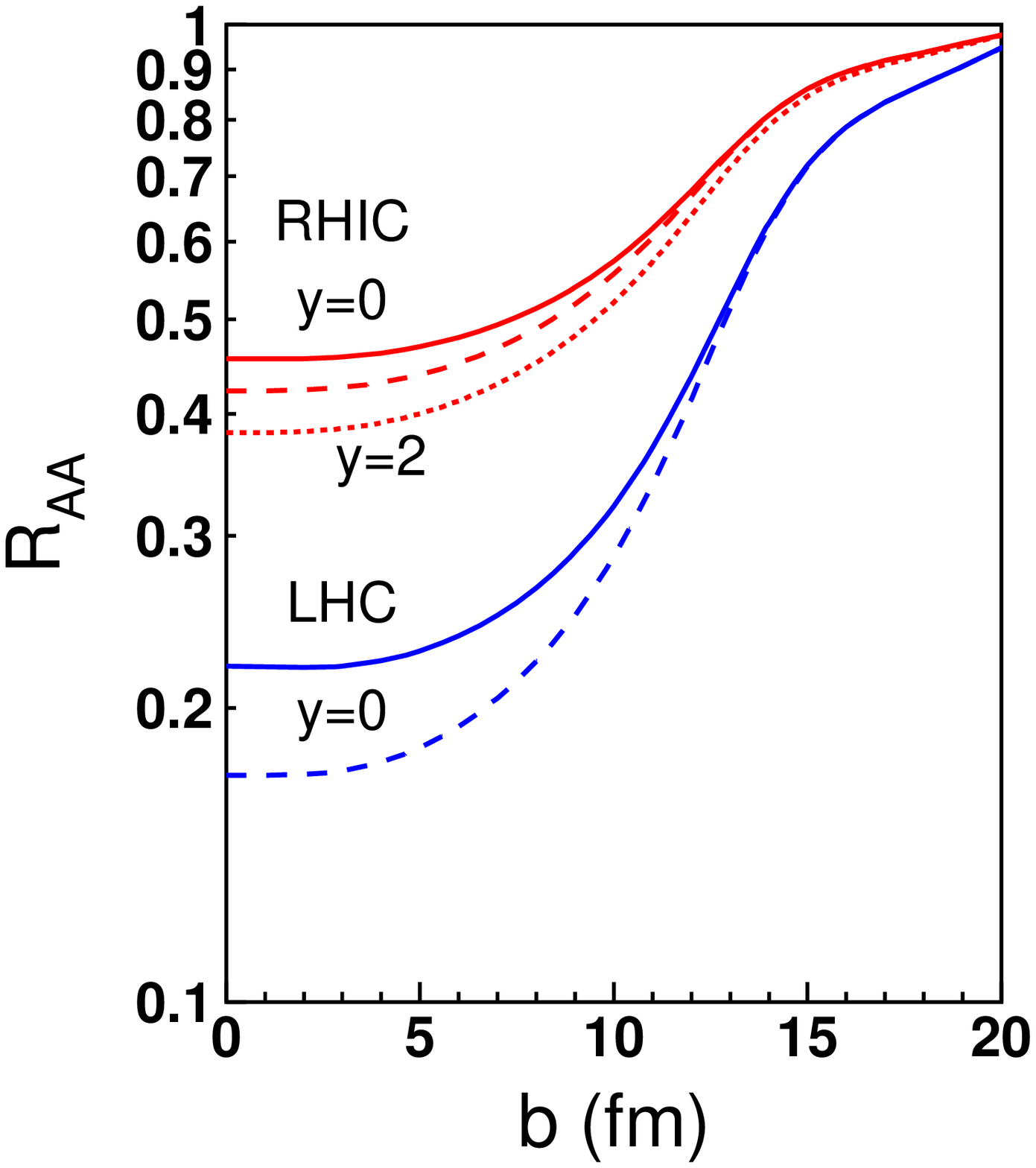}}}
\caption{{\it Left:} The nuclear ratio $R_{AA}$ for central ($b=0$) collisions of gold-gold (RHIC) and lead-lead (LHC) vs impact parameter $\tau$. The upper and bottom dashed curves correspond to calculations with the same saturation scale as in $pA$ at $\sqrt{s}=200\GeV,\ 5.5\TeV$ respectively. Solid curves are calculated with the boosted saturation scale, which makes the nuclei more opaque.
{\it Right:} The $\tau$-integrated $R_{AA}(b)$ for $J/\Psi$ suppression by the ISI. 
Solid and dashed curves present the results at $y=0$ including  and excluding the effect of double-color-filtering, respectively. The upper and bottom pairs of curves (solid and dashed) correspond to $y=0$ and energies $\sqrt{s}=200\GeV$ and $5.5\TeV$ respectively. 
The dotted curve is calculated at $y=2$.
}
\label{fig:boosting}      
\end{figure}

Solving these equations one can calculate the ISI suppression of $\J$ in $AA$ collisions and compare with the results obtained with "normal" value of $Q_s$, the same as in $pA$ collisions. This comparison presented in Fig.~\ref{fig:boosting} (left). 
While at the energy of RHIC the boosting effect is rather mild, in the LHC energies the boosted saturation scale makes the nuclear medium significantly more opaque, and $\J$ is several times more suppressed compared to the simplified expectations.

The  rise of the saturation scale means an increase of broadening \cite{broadening} in $AA$ versus $pA$ collisions. This effect looks similar to what was observed at SPS as we discussed in the previous Sect.~\ref{broad}.
However that one disappears at the energies of RHIC and LHC, while the boosted saturation scale is a coherence effect, which sets on only at high energies. 

Thus, we again conclude that the so called ``cold nuclear matter" in $AA$ collisions is not cold.

\subsection{Double color filtering}

In a Glauber-like approach the survival probability of a $\bar cc$ dipole propagating through the colliding nuclei at the ISI stage is a product of the survival probabilities in each of the nuclei. This also seems to go along with the conventional intuition.
We demonstrate, however, that at high energies this is not correct, and the double-color-filtering effect makes the transition from $pA$ to $AA$ not so straightforward, as usually believed \cite{nontrivial}.

As an example, let us take a piece of nuclear matter of constant density $\rho$. Classically, the survival probability of a dipole propagating a path length $L$ in the medium is $P(L)=e^{-\sigma_{abs}\,\rho L}$ (compare with (\ref{10})). However, the absorption cross section is an average over the dipole size weighted with the dipole wave function,
$\sigma_{abs}=\la\sigma_{\bar QQ}(r_T)\ra= \int d^2r_T\,|\Psi_{\bar qq}(r_T)|^2\sigma_{\bar qq}(r_T)$. At high energies $l_f\gg L$ the whole exponential must be averaged, rather than just the exponent (like in the Glauber model),
\beq
P(L)=\left\la e^{-\sigma_{\bar qq}(r_T)\,\rho L}\right\ra=
\frac{1}{1+\sigma_{abs}\,\rho L}
\label{150}
\eeq
For the sake of simplicity we assumed  a gaussian form for $\Psi_{\bar qq}(r_T)$, and $\sigma_{\bar qq}(r_T)=C\,r_T^2$.
The result shows that the effect of color transparency makes the medium more transparent (as expected).

Naively, one could guess that the survival probability for simultaneous propagation through the two nuclei with the path lengths $L_A$ and $L_B$ has the form of a simple product,
\beq
P(L_A,L_B)=P(L_A)P(L_B)=
\frac{1}{(1+\sigma_{abs}\,\rho L_A)(1+\sigma_{abs}\,\rho L_B)}\ .
\label{160}
\eeq
However, in this case one should average over $r_T$ the product of two exponentials with the result,
\beq
P(L_A,L_B)=\frac{1}{1+\sigma_{abs}\,\rho(L_A+L_B)},
\label{170}
\eeq
which is quite different from (\ref{160}). This difference comes from the effect of double color filtering, which makes the medium more transparent because filtering in one nucleus reduces the average size of the survived dipoles, so that the other nucleus becomes more transparent.

\begin{figure}[h!]
\centerline{
\resizebox{0.41\columnwidth}{!}{
\includegraphics{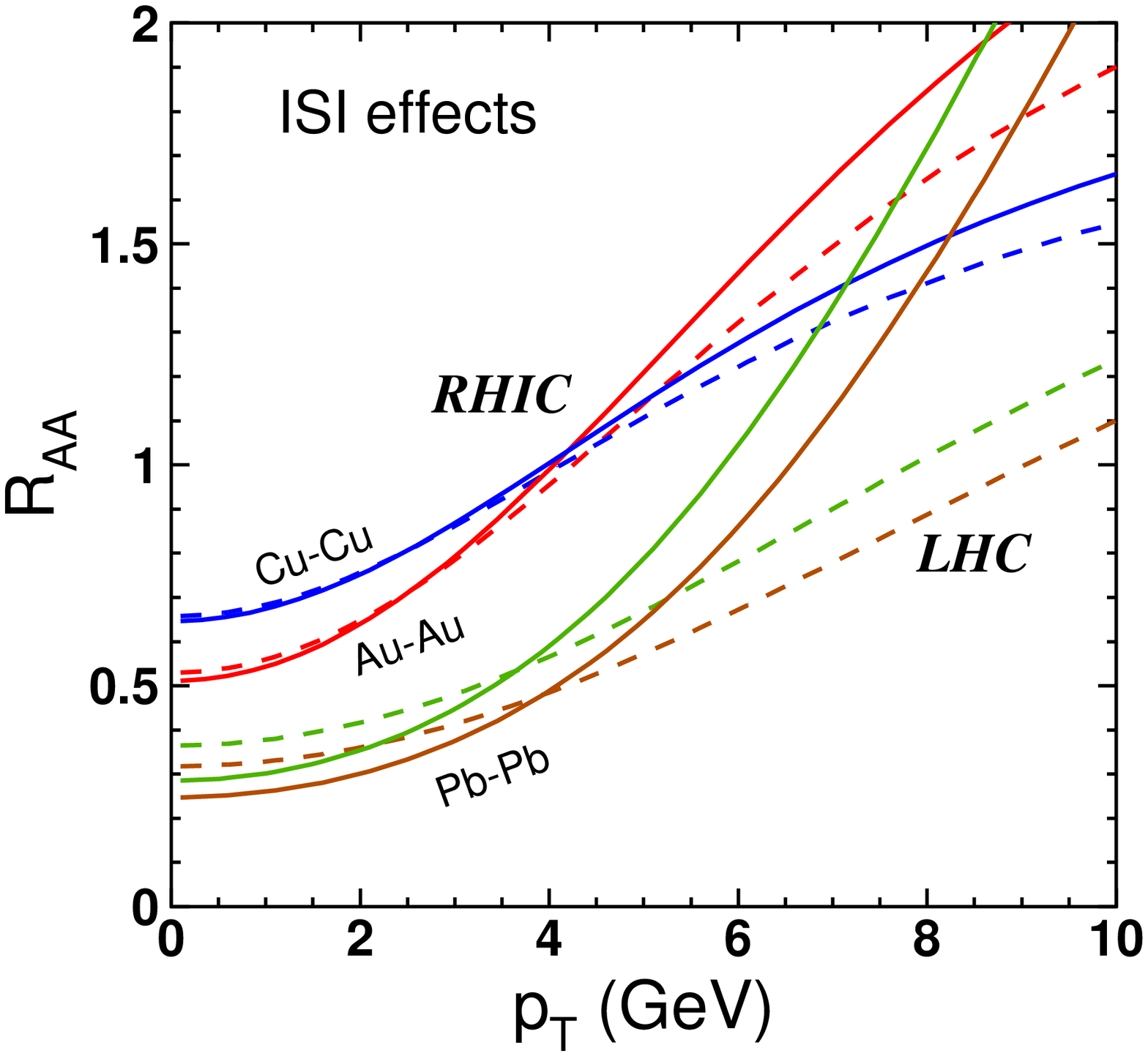} }\hspace{1cm}
\resizebox{0.4\columnwidth}{!}{
\includegraphics{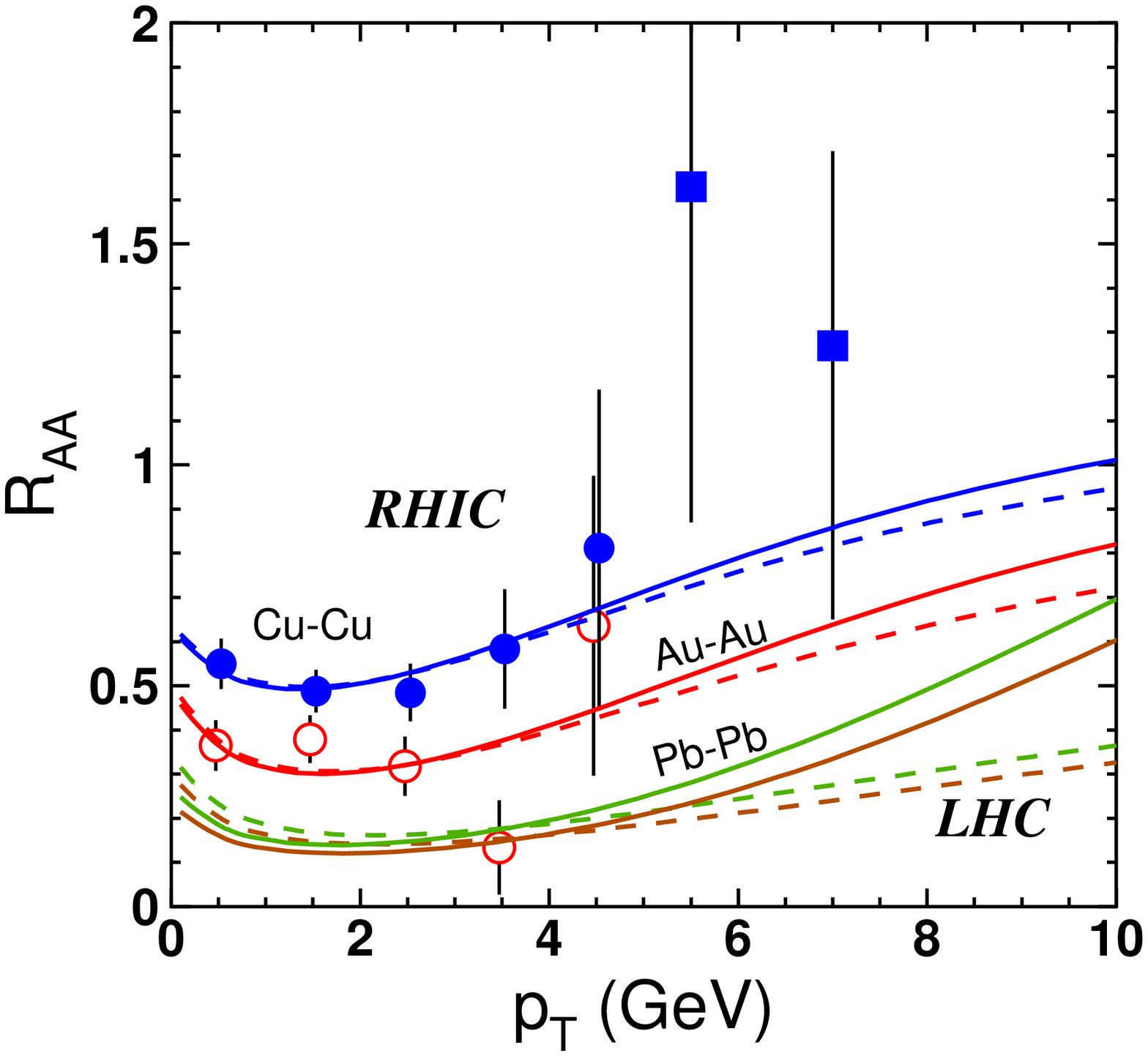}}}
\caption{{\it Left:} Nuclear ratio for $\J$ produced in central ($b=0$) $AA$ collisions including only the ISI effects. Dashed lines correspond to (from top to bottom) $Cu$-$Cu$ and $Au$-$Au$ at $\sqrt{s}=200\GeV$, $Pb$-$Pb$ at $\sqrt{s}=2.76$ and $5.5\TeV$, respectively. The solid curves include the effects of boosted saturation scale and double color filtering.
{\it Right:} RHIC data \cite{phenix1,phenix2,star} for $\J$ production in central collisions of $Cu$-$Cu$ (closed circles and squares) and $Au$-$Au$ (open circles) at $\sqrt{s}=200\GeV$ and $y=0$.
The curves are the same as in the left pane of this figure, but corrected for the FSI effects, calculated with the transport coefficient parameter for gold-gold $\hat q_0=0.6\GeV^2/\fm$ adjusted to the data in  \cite{psi-bnl}. $R_{AA}$ at the LHC energies is predicted with $\hat q_0=0.8\GeV^2/\fm$, extracted in \cite{high-pt} from data for nuclear quenching of high-$p_T$ hadrons \cite{alice} (see text). } 
\label{fig:fsi}      
\end{figure}

Numerically, this effect is not very strong, as is demonstrated in Fig.~\ref{fig:boosting} (right). It works in the opposite direction to the boosting effect, which is numerically stronger and makes the colliding nuclei more opaque.
The combined effect of both phenomena on the $\J$ suppression at the ISI stage (including also the Cronin effect) is shown by solid curves in Fig.~\ref{fig:fsi} (left), 
We see that the boosting effect significantly enhances the $\J$ production rate at large $p_T$ at LHC. This happens because broadening is boosted.

\section{AA collisions: combining the ISI and FSI effects}

Considering $\J$ produced with  where 
Data from RHIC for the $p_T$ dependence of $\J$ production with a reasonable accuracy  are currently available
at $p_T\lesssim 5\GeV$. In this kinematical domain one can evaluate the characteristic time scales. 
The production and formation times  in the rest frame of the medium are given by,
\beqn
t_p^*&=&\frac{1}{\sqrt{4m_c^2+p_T^2}} < 0.07\fm
\label{165}\\
t_f^*&=&\frac{\sqrt{p_T^2+M_{\J}^2}}{(M_{\Psi'}-M_{\J})M_{\psi}}\lesssim 0.5\fm.
\label{170}
\eeqn
are shorter than the time scale of medium creation, $t_0\sim1\fm$, and much shorter than the mean path length in the medium $L\sim 5\fm$.           
So we conclude that differently from the high energy limit, where a ``frozen" $\bar cc$ dipole propagates through the final state medium, in this case it is a fully formed   $J/\Psi$, so one can rely on the Glauber approximation.

The dipole cross section is related by Eq.~(\ref{110}) to parton broadening, i.e. to the transport coefficient $\hat q$, which is defined as the broadening rate per unit of length,
$\hat q=2\,\rho\,d\sigma(r_T)/dr_T^2|_{r=0}$.  Thus, the dipole break-up rate is also related to $\hat q$ as
$dS(r_T,l)/dl = -{1\over2}\,\hat q\,r_T^2$ \cite{psi-AA,psi-bnl}. The FSI modification factor gets the form,
\beq
R(s,p_T)=
\frac{1}{\pi}\int\limits_0^\pi d\phi\,
\exp\Biggl[-\,\frac{1}{2}\,\la r_{\psi}^2\ra
\int\limits_{l_0}^\infty dl\ \hat q(\vec{\bf s}+\vec{\bf l})
\Biggr]
\label{180}
\eeq

For the coordinate and time dependence of the transport coefficient one can employ the popular model \cite{psi-AA,psi-bnl}
\beq
\hat q(\vec b,\vec\tau,t)=\frac{\hat q_0\,t_0}{t}\,\frac{n_{part}(\vec b,\vec\tau)}{n_{part}(0,0)}
\label{190}
\eeq

The magnitude of the transport coefficient was adjusted  in \cite{psi-AA,psi-bnl} at $\hat q_0=0.6\GeV^2/\fm$  to reproduce the data, as is depicted in Fig.~\ref{fig:fsi} (right). The ISI effects are included. The smallness of $\hat q_0$ resolves the puzzle of the unexpectedly large (20 times larger) value extracted from jet quenching data within the energy loss scenrio~\cite{phenix-theor}.

The nuclear modification factor at the energies of LHC was also predicted in \cite{psi-lhc}
relying on the result $\hat q_0=0.8\GeV^2/\fm$ of the analysis \cite{high-pt} of ALICE data for high-$p_T$ hadron suppression \cite{alice} performed within a scenario based on dipole attenuation .
Notice than the recent more advanced analysis \cite{path} of RHIC and LHC data for high-$p_T$ hadrons employing  the path-integral technique, led to close, but somewhat higher values $\hat q_0=1.6$ and $2\GeV^2/\fm$ respectively. The difference is probably related to the approximations made above. In particular,
the quadratic rise of the dipole cross section, $\sigma(r_T)\propto r_t^2$ up to a size of $\J$ certainly overestimates the absorption cross section in Eq.~(\ref{180}). Also the formation time Eq.~(\ref{170}) was completely neglected, but it would have been more accurate to use the path integral technique. Both this approximations lead to to a more opaque medium, i.e. to a diminished value of $\hat q_0$ adjusted to data. A more comprehensive analysis of $\J$ suppression is in progress.
 
Another potential source  of a missed dynamics, which may lead to a reduction of the effective $\hat q_0$
is the thermal mechanism of $\J$ production \cite{kabana,peter-jochanna,scanning} due to coalescence of charm quarks abundantly produced at high energies. Although, no clear signal of this mechanism has been ob served in the experiments at the SPS and RHIC, it seems to be the only way to understand the unusual behavior of the nuclear effects as function of $p_T$ and centrality observed at the LHC \cite{qm2012}. A more detailed discussion of this mechanism is beyond the scopes of this paper.

\section{Summary}

\begin{itemize}

\item
Interplay between the effects of color transparency in the final state and charm shadowing in the initial state
leads to a peculiar energy dependence of the effective absorption cross section of $\J$ produced in $pA$ collisions. It rises up to some energy, as is confirmed by fixed target experiments, then it starts falling with energy being suppressed by charm shadowing. 

\item
At the energies of RHIC and LHC both effects reach the asymptotic regime and become the main source of $\J$ suppression, although they are higher twists. The leading twist gluon shadowing is found rather weak.

\item
The transition from $pA$ to $AA$ collisions is not as straightforward as is usually believed. 
The produced $\bar cc$ dipole interacts not with bound nucleons, but with the colored debris of those who have already had a chance to interact prior meeting the charm dipole. This leads to an increase break-up cross section of the dipole, as well
to an enhanced broadening of the primordial projectile gluon. Both effects, anomalous (compared with $pA$) $\J$ suppression and broadening were observed in $AA$ collisions at SPS, and both vanish at higher energies of RHIC and LHC.

\item
New effect of a boosted saturation scales affecting ISI in colliding nuclei onsets at high energies. Although this effect has a different origin, it acts similar to what was observed at the energies of SPS, namely the increase of the saturation scale leads to a 
stronger suppression of $\J$ by ISI and to a larger
broadening. While the magnitude of the effect is rather mild at RHIC, it is grossly enhances at the energies of LHC.

\item
The affect of double color filtering acts in the opposite direction to the boosting, it makes the nuclei more transparent.

\item
$J/\Psi$ production offers an alternative probe for the transport coefficient of the medium created in heavy ion collisions.
The final state attenuation of $\J$ is controlled by the same transport coefficient as parton broadening and energy loss. The small transport coefficient  found from  the analysis of  $J/\Psi$ data is close to what has been predicted, but is much smaller than the result of jet quenching analyses based on the energy loss scenario.

\end{itemize}

{\bf Acknowledgements:} B.Z.K. thanks the organizers of the International Conference on New Frontiers in Physics for the invitation to deliver  this talk. This work was supported in part by Fondecyt (Chile)
grants 1090291, 1090236 and 1100287.

\end{document}